\documentclass[fleqn,usenatbib]{mnras}

\usepackage{newtxtext,newtxmath}

\usepackage[T1]{fontenc}
\usepackage{color,soul}

\DeclareRobustCommand{\VAN}[3]{#2}
\let\VANthebibliography\thebibliography
\def\thebibliography{\DeclareRobustCommand{\VAN}[3]{##3}\VANthebibliography}

\usepackage{graphicx}	
\usepackage{amsmath}	

\title[Dark acoustic oscillations]{Dark acoustic oscillations: Imprints on the matter power spectrum and the halo mass function}
\author[T. Schaeffer et al.]{
Timoth\'ee Schaeffer$^{1}$\thanks{E-mail: timothee.schaeffer@uzh.ch} and Aurel Schneider$^{1}$\thanks{E-mail: aurel.schneider@uzh.ch}
\\
$^{1}$Center for Theoretical Astrophysics and Cosmology, Institute for Computational Science, University of Zurich, \\Winterthurerstrasse 190, CH-8057 Z{\"u}rich, Switzerland\\
}

\date{Accepted XXX. Received YYY; in original form ZZZ}

\pubyear{2021}

\begin{document}
\label{firstpage}
\pagerange{\pageref{firstpage}--\pageref{lastpage}}
\maketitle

\begin{abstract}
Many non-minimal dark matter scenarios lead to oscillatory features in the matter power spectrum induced by interactions either within the dark sector or with particles from the standard model. Observing such dark acoustic oscillations would therefore be a major step towards understanding dark matter. We investigate what happens to oscillatory features during the process of nonlinear structure formation. We show that at the level of the power spectrum, oscillations are smoothed out by nonlinear mode coupling, gradually disappearing towards lower redshifts. In the halo mass function, however, the oscillatory features remain visible until the present epoch. As a consequence, dark acoustic oscillations could be detectable in observations that are either based on the halo mass function or on the high-redshift power spectrum. We investigate the  effect of such features on  different observables, namely, the cluster mass function, the stellar-to-halo mass relation, and the Lyman-$\alpha$ flux power spectrum. We find that oscillatory features remain visible in all of these observables, but they are very extended and of low amplitude, making it challenging to detect them as distinct features in the data.

\end{abstract}




\section{Introduction}

In the standard cosmological model of $\Lambda$CDM, dark matter (DM) is described as a cold and collisionless fluid, consisting of one fundamental particle with negligible interaction properties. While these assumptions provide a minimal working setup for cosmology, the dark matter sector could in principle be considerably more complex. For example, it is possible that dark matter interacts with itself (via a dark mediator) or with other species of the standard model. Furthermore, the DM fluid could consist of more than one particle, making room for a plethora of dark interactions, decays, or annihilation processes.

A rather generic outcome of interactions involving dark matter is the emergence of dark acoustic oscillations (DAO). Such oscillatory features can arise due to similar physical processes than the ones at the origin of the baryon acoustic oscillations (BAO), i.e. interactions in the dark matter sector, forming pressure-gravity waves that are imprinted on the clustering signal. Dark acoustic oscillations have been predicted for scenarios of DM-photon \citep{DM_photons_Boehm} and DM-neutrino interactions \citep{DM_neutrino_int_frac,DM-nu_Boehm}, models including dark radiation \citep{Buckeley_DM_DR,Dissipative_DM_Sunny,Foot_2016,AssymetricDM,Ballistic_DM}, dark atom scenarios \citep{Dark_atom_model}, or through alternative scenarios of inflation \citep{Lensing_anomaly_inflation_osc,Inflation_osc_summary}.

Most dark acoustic oscillations predicted by the models mentioned above, tend to appear at the small, highly-nonlinear clustering scales, often in combination with a strong damping effect. The reason for this behaviour is that, unlike the BAO features that appear at quasi-linear scales, the oscillations are driven by the DM fluid itself, which dominates the matter budget of the universe. However, smaller dark acoustic oscillations at larger scales and without strong damping of the power spectrum may naturally appear if only a fraction of the dark matter budget is involved in the interaction process.

Regardless of whether parts or all of dark matter is interacting, it is currently unclear if and for how long oscillatory features remain detectable in the nonlinear regime of structure formation. We know that nonlinear mode-coupling effects tend to smooth out small-scale features in the power spectrum \citep{NonLinear_Regime_BAO,NonLinear_Oscillations_Power} but it remains unclear how fast this smoothing proceeds. Furthermore, the oscillations could remain visible in other statistics of the large scale structure such as, for example, the distribution of haloes.

In this paper, we study the evolution of DM-induced oscillation patterns in the matter power spectrum and the halo mass function. We thereby rely on the ETHOS scenario \citep{Ethos_eff}, which consists of an effective two-component model including dark matter and dark radiation. Additionally to the original ETHOS framework, we furthermore allow for another non-interacting dark matter component, acting as a cold and collisionless fluid. This allows us to not only investigate the case of damped oscillations at very small scales, but also more subtle oscillatory features that may appear at a wide range of different cosmological scales.

We investigate three main scenarios: a subdominant interacting sector leading to small oscillatory features at wave modes $k\sim 0.1-1$ h/Mpc, a strongly dominating interacting sector resulting in large oscillations at $k\sim 10-100$ h/Mpc, and an intermediate scenario featuring medium sized oscillations at scales $k\sim 1-10$ h/Mpc. Note that all these scales are in the non-linear regime of structure formation today.

Next to studying the fate of oscillations in the matter power spectrum and the halo mass function, we also discuss the prospects of a potential future detection. In particular, we investigate the cluster mass function, the stellar-to-halo mass relation, and the Lymna-$\alpha$ forest as potential observables. We thereby specifically focus on the detectability of the oscillation features and not so much on the additional power suppression that typically manifests itself at similar wave modes. While generic suppression effects of the power spectrum can be caused by many phenomena related to cosmology, oscillations features are a smoking-gun for dark matter particle-interactions in the early universe and are unlikely to be degenerate with other signals from cosmology or astrophysics.

The paper is structured as follows. In Sec.~\ref{sec2} we discuss the interacting dark matter framework and introduce our suite of numerical simulations. Sec.~\ref{sec3} and \ref{sec4} are dedicated to the measurements of the matter power spectrum and halo mass function with a specific focus on the oscillation features. In Sec.~\ref{sec5} we discuss different observables and their potential to detect oscillations before concluding in Sec.~\ref{sec6}.

\section{Interacting dark matter}\label{sec2}
There are many ways dark matter (DM) may interact with other dark components or with particles of the standard model. In this section we define the effective DM framework used in this paper, and we discus the setup for the calculations of the linear and nonlinear perturbations.

\subsection{Framework of the model}
As mentioned in the introduction, several distinct DM scenarios may lead to oscillation patterns in the linear matter power spectrum. While we will attempt to be agnostic about the exact model in our discussions, we are forced to select a specific model for the calculations. This means that at the quantitative level, our results are only strictly accurate for our specific setup. However, we expect the main findings to be more generally valid and relatable to other DM models featuring generic oscillations.

We assume the total dark matter to be made of a cold and non-interacting (collisionless) dark matter component (CDM) plus an interacting sector consisting of a dark matter (IDM) and dark radiation component (IDR). The fraction of interacting to total DM is given by 
\begin{equation}
    f_{\rm IDM} = \frac{\Omega_{\rm IDM}}{\Omega_{\rm IDM}+\Omega_{\rm CDM}},
\end{equation}
where $\Omega_{\rm IDM}$ and $\Omega_{\rm CDM}$ denote the abundances of the interacting and non-interacting DM components. The interacting DM sector is described by the ETHOS parametrisation presented in \cite{Ethos_eff}.

The ETHOS framework introduces a set of parameters describing the effects of generic interactions between the IDM and the IDR mediator particles at the level of the Boltzmann equations. The model parameters originate from an expansion of the IDM and IDR opacity coefficients $\Dot{\kappa}$ in powers of temperature. The corresponding coefficients $a_{n}$ and $b_{n}$, the power $n$ of the expansion, along with the amount of dark radiation (defined by the temperature ratio $\xi=(T_{\mathrm{DR}}/T_{\mathrm{CMB}})|_{\mathrm{z=0}}$) act as the free model parameters. In terms of the resulting cosmological perturbations, they determine the scales where dark acoustic oscillations appear and the degree of damping affecting the small-scale wave modes. 

As explained in details in \cite{JeanScaleEthos}, the evolution of the DM-DR coupling in ETHOS can be separated in three main stages: (i) the \textit{tight coupling epoch} ($\dot{\kappa}_{\mathrm{DM-DR}}\gg aH$), where sound waves propagate due to the effective radiative pressure in the dark plasma, giving rise to dark acoustic oscillations; (ii) the \textit{drag epoch} ($\dot{\kappa}_{\mathrm{DM-DR}}\sim aH$), where the dark radiation decouples and starts to free stream, while the DM is still kinematically coupled to the DR bath; and (iii) the \textit{decoupled epoch} ($\dot{\kappa}_{\mathrm{DM}}\ll aH$), where interactions become negligible and the DM perturbations grow in the standard way. Note that in terms of the DAO, the shorter the drag epoch, the less damped the first acoustic peaks, and the more visible the oscillations in the matter power spectrum. Fast transitions from the coupled to the uncoupled regime require a steep function of $\dot\kappa(T)$, which is obtained by assuming high values of the expansion power $n$. For instance, certain configurations of the dark atom model (\citealt{Dark_atom_model}) featuring a fast dark recombination can be mimicked with the Ethos parametrization by picking a large value of n.

All results shown in this paper are based on cosmological parameters from the Planck18 survey \citep{Planck2018}. In particular we use $\Omega_{m}=0.321$, $\Omega_{b}=0.049$, $\Omega_{\Lambda}=0.679$, $H_{0}=66.88$, and $n_s=0.96$.



\subsection{Linear perturbations}
The linear matter perturbations of interacting DM are obtained by solving the coupled Boltzmann equations from ETHOS. A detailed discussion of this calculation for several different parameter choices has been presented in the original ETHOS paper \citep{Ethos_eff}. The same authors also provide a code based on the publicly available Boltzmann solver {\tt CAMB} (\citealt{CAMB_Lewis_2000,Ethos_eff}).

More recently, the ETHOS framework has been implemented into the Boltzmann code {\tt CLASS} (\citealt{Class_code,Class_Ethos}).
{\tt CLASS} allows for multiple DM scenarios, and is more robust than the original ETHOS code regarding numerical instabilities. For this reason, we use {\tt CLASS} for all calculations in the present paper.

\begin{figure*}
     \centering
     \includegraphics[width =1\textwidth,trim=3.5cm 0.cm 4.cm 1.15cm,clip]{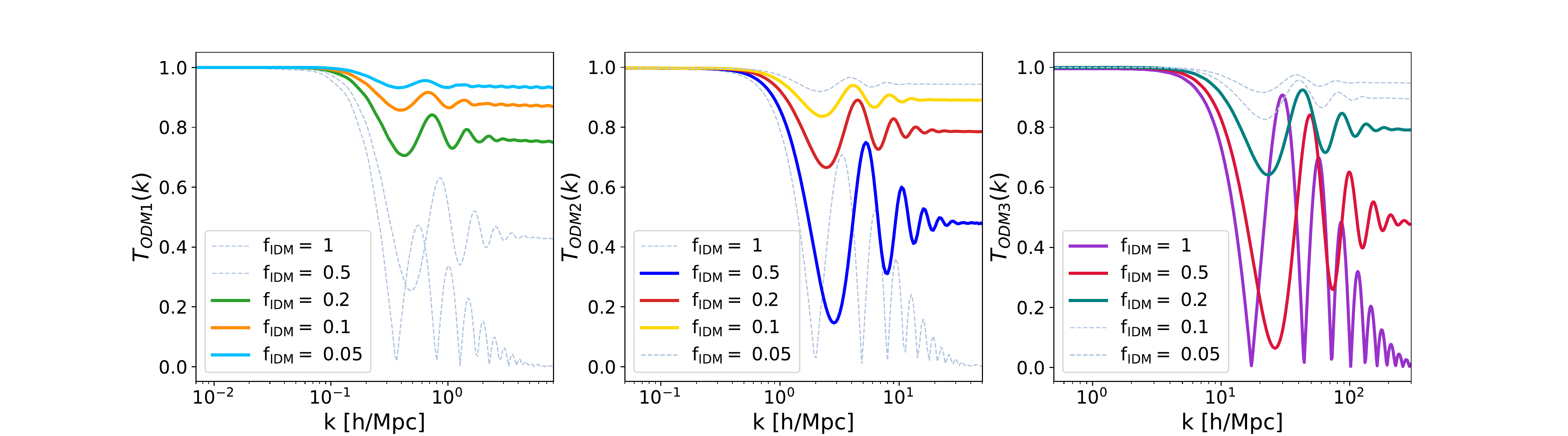}
    \caption{Initial transfer functions (defined in Eq.~\ref{eq:T}) for the ODM1, ODM2 and ODM3 models with interacting-to-total DM fractions $f_{\rm IDM}=0.05$, 0.1, 0.2, 0.5 and 1. All models that we investigate further by running $N$-body simulations are highlighted with coloured lines.}
     \label{fig:T-K_ODM}
\end{figure*}

Since our goal is to study how primordial oscillations evolve through non-linear gravitational clustering, we choose models that minimize the damping of the dark acoustic oscillations. This is achieved by fixing $n=6$ which leads to a quasi-instantaneous decoupling transition, i.e a small drag scale \citep[see appendix A.1 of][]{JeanScaleEthos}. We then vary the parameters $a_{6}$, $\xi$, and $f_{\rm IDM}$ in order to obtain various models of oscillating dark matter (ODM) as summarised in table \ref{tab:DM models param}. The models differ by the range of scales where oscillations appear in the matter power spectrum as well as by the amplitude of the oscillatory features. We thereby categorise three main classes of models (ODM1, ODM2, and ODM3) with oscillations appearing at $k=0.1-1$ h/Mpc, $k=1-10$ h/Mpc, and $k=10-100$ h/Mpc. These wave modes correspond to the galaxy cluster, the Milky-Way type galaxy, and the dwarf galaxy scale, respectively. For each class of model, we furthermore vary the fraction of interacting to total DM with $f_{\rm IDM}=\{0.05, 0.1, 0.2\}$ for ODM1, $f_{\rm IDM}=\{0.1, 0.2, 0.5\}$ for ODM2, and $f_{\rm IDM}=\{0.2, 0.5, 1.0\}$ for ODM3.

\begin{table}
 \begin{center}
  \setlength{\tabcolsep}{1.7\tabcolsep}
  \begin{tabular}{ccccc}
  \hline
  Model & $f_{\rm IDM}$ & $a_{6}$ & $\xi$ \\ \hline
  ODM1 & 0.05, 0.1, 0.2  & 1e18 & 0.5   \\
  ODM2 & 0.1, 0.2, 0.5  & 1e14 & 0.3    \\
  ODM3 & 0.2, 0.5, 1  & 6e8  & 0.184  \\
  \hline
  \label{tab:DM models param}
  \end{tabular}
  \caption{ETHOS Parametrisation for the dark matter models studied in this paper. The power-law scaling of the DM-DR scattering rate with temperature is n=6. We vary the coefficient $a_{6}$, the amount of dark radiation via $\xi$ and the amount of interacting DM via $f_{\rm IDM}$. Note that the IDM mass is kept to 1 GeV, and that other parameters such as the dark interactions angular coefficients are kept as default ones.}
 \end{center}
\end{table}

In Fig.~\ref{fig:T-K_ODM} we show the linear transfer functions ($T_{\rm ODMi}$) for several different ETHOS models with $n=6$ and varying $a_{6}$, $\xi$, and $f_{\rm IDM}$. The three panels refer to the ODM1, ODM2, and ODM3 classes of models while coloured lines represent the different DM fractions listed in table \ref{tab:DM models param}. Note that the transfer function is defined as the square-root of the ratios of ODM to CDM linear power spectra, i.e.
\begin{equation}
    T_{\rm ODMi}=\sqrt{P_{\rm ODMi}/P_{\rm CDM}},\hspace{0.5cm} {\rm for}\,\, i=1,2,3.
    \label{eq:T}
\end{equation}
All transfer functions plotted in Fig.~\ref{fig:T-K_ODM} are exactly one for very low $k$-values (large physical scales), before showing characteristic oscillations plus an overall decrease of power towards smaller scales. At very high values of $k$, all curves reach a plateau with an amplitude depending solely on $f_{\rm IDM}$. The oscillatory features exhibit a higher amplitude for larger values of $f_{\rm IDM}$ and if they appear at smaller scales. However, they remain clearly visible even for cases where less than ten percent of the DM budget is interacting. These models with small fractions of interacting DM are particularly interesting as they remain largely unconstrained by current observations.

\begin{figure*}
     \centering
     \includegraphics[width =1\textwidth,trim=0cm 0.2cm 0cm 0.cm,clip]{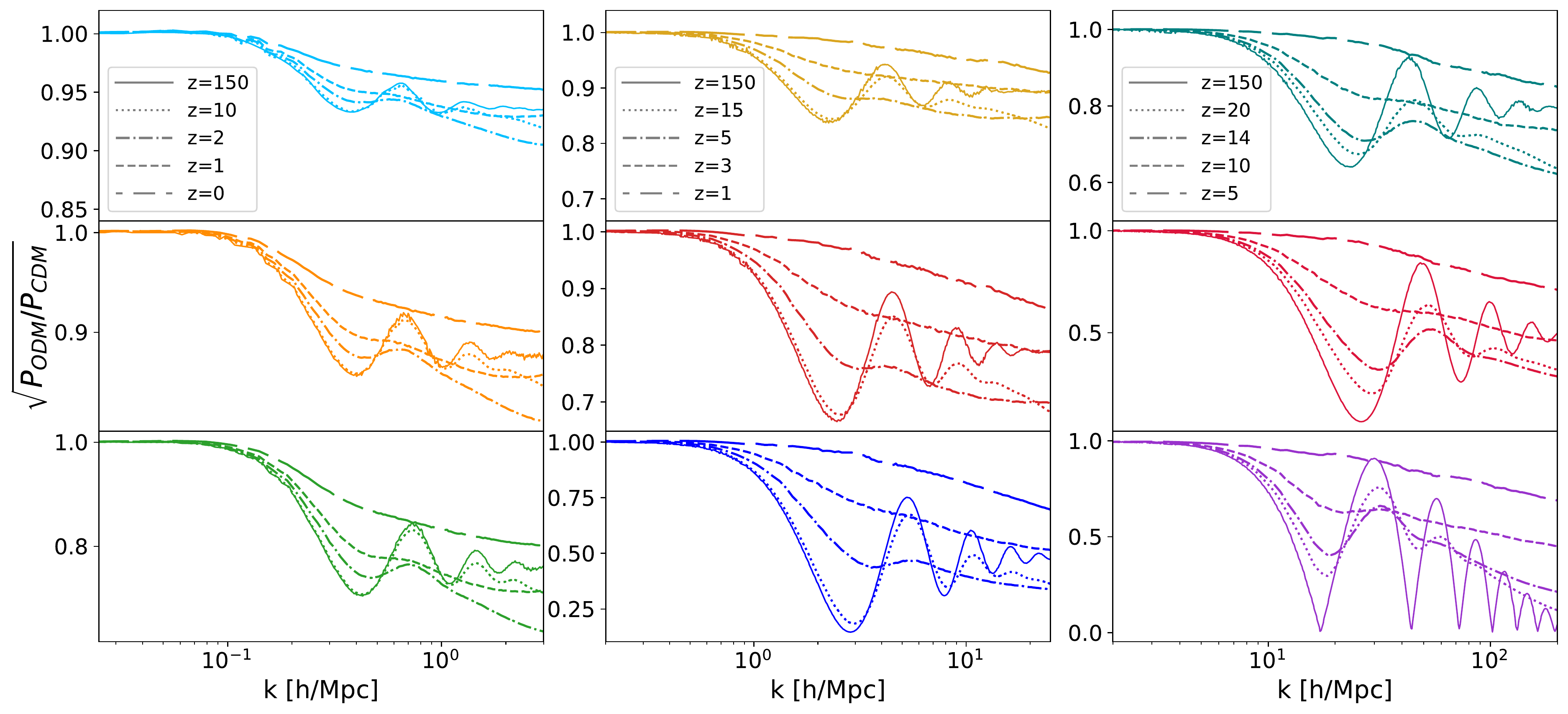}
    \caption{Square roots of the ratios between the ODM and CDM nonlinear power spectra at different redshifts. Solid lines correspond to the initial transfer functions from Fig .\ref{fig:T-K_ODM}. \textit{Left panels}: ODM1 at $z=10, 2, 1, 0$ (from top to bottom: $f_{\rm IDM} = 0.05, 0.1 ,0.2$), ODM2 at $z=15,5,2,0$ ($f_{\rm IDM} = 0.1, 0.2 ,0.5$), and ODM3 at $z=20,14,10,3.5$ ($f_{\rm IDM} = 0.2, 0.5 ,1$). All oscillations have completely disappeared at $z=1$ for ODM1, $z=3$ for ODM2, and $z=10$ for ODM3.}
     \label{fig:Non_linear_T-K_ODM}
\end{figure*}

We would like to stress that the transfer function of all ODM models show two main features: (i) an overall suppression effect towards smaller scales and (ii) an oscillation signal with multiple peaks. Both of these features are expected to contribute to distinguishing ODM models from CDM, using current or future observations. However, while (i) may be largely degenerate with many other cosmological or astrophysical effects, (ii) consists of a characteristic signal that may act as a smoking gun for interactions in the DM sector. This is why we are particularly interested in studying the evolution of oscillatory features during nonlinear structure formation.

\subsection{Numerical Simulations}
We use gravity-only $N$-body simulations to study the non-linear structure formation of the oscillating DM models introduced above. All simulations are run with {\tt Pkdgrav3}, a tree-code based on the fast multipole method and using adaptive time-stepping \citep{stadel2001,pkdgrav3}. The initial conditions are set up assuming 2LPT and using identical random seeds to allow for direct comparisons between simulations without having to worry about cosmic variance.

\begin{table}
 \begin{center}
  \setlength{\tabcolsep}{1.7\tabcolsep}
  \begin{tabular}{cccccc}
  \hline
  Model & $L_{\rm{box}}$ [Mpc/h] & $N_{p}$ &  $m_{p}$ [M$_{\odot}$/h] & $z_{0}$  \\ \hline
  ODM1: & 2048 & 512$^3$ & $5.7\times10^{12}$ &  0 \\
   &  2048  & 1024$^3$ & $7.1\times10^{11}$  &  0  \\
   &  512 & 512$^3$ & $8.9\times10^{10}$   &  0  \\
   &  512 & 1024$^3$ & $1.1\times10^{10}$  &  0 \\
   \hline
  ODM2: &  512 & 512$^3$  & $8.9\times10^{10}$  &  0 \\
   &  512 & 1024$^3$ & $1.1\times10^{10}$, &  0 \\
   &  64 &  512$^3$ &  $1.7\times10^{8}$  &  0  \\
   &  64 & 1024$^3$ &  $2.2\times10^{7}$  &  0  \\
   \hline
  ODM3: & 32 & 512$^3$ &  $2.2\times10^{7}$  &  2 \\
    &  32 & 1024$^3$ &  $2.7\times10^{6}$  &  2 \\
    &  8 & 512$^3$ &  $3.4\times10^{5}$  &  2 \\
    &  8& 1024$^3$  & $4.2\times10^{4}$  &  2    \\
  \hline
  \label{tab:SimulationSetup}
  \end{tabular}
  \caption{Details of the simulation suite with specifications about the box size ($L_{\mathrm{box}}$), the particle number ($N_{p}$), the corresponding mass of one individual simulation particle ($m_{p}$) and the final redshift ($z_0$) of the run.}
  \end{center}
\end{table}

A summary of all $N$-body runs is provided in Table~\ref{tab:SimulationSetup}. The box length ($L_{\rm box}$) and particle numbers ($N_{p}$) of the simulations are selected so that all main oscillatory features are well resolved. For the ODM1 model we use $L_{\mathrm{box}}=512$ Mpc/h  and $L_{\mathrm{box}}=2048$ Mpc/h, for ODM2 we use $L_{\mathrm{box}}=64$ Mpc/h and $L_{\mathrm{box}}=512$ Mpc/h, and for ODM3  $L_{\mathrm{box}}=8$ Mpc/h and $L_{\mathrm{box}}=32$ Mpc/h. All simulations are run with both $N_p=512^3$ and $N_p=1024^3$ particles. This allows us to check for resolution issues that could affect the results (see appendix \ref{appendixSpurious} about artificial haloes).

The $N$-body simulations are all started at the initial redshift $z_i=150$. While we run the ODM1 and ODM2 models to $z_f=0$, the ODM3 simulations stop at $z_f=2$. This ensures the linearity of modes at the scale of the simulation box size.

With the simulation suite described above, we are able obtain fully converged nonlinear matter power spectra and halo mass functions at the relevant scales where the dark acoustic oscillations occur. Note that individual haloes can be resolved with at least 500 particles at scales around the main oscillation peak. Only such large numbers make it possible to reliably exclude the presence of artificial haloes which are known to populate simulations of both warm and mixed DM models \citep{AurelSupressedSmallScales}.

\begin{figure*}
    \centering
    \includegraphics[width =1\textwidth,trim=1.5cm 0.2cm 2cm 1.15cm,clip]{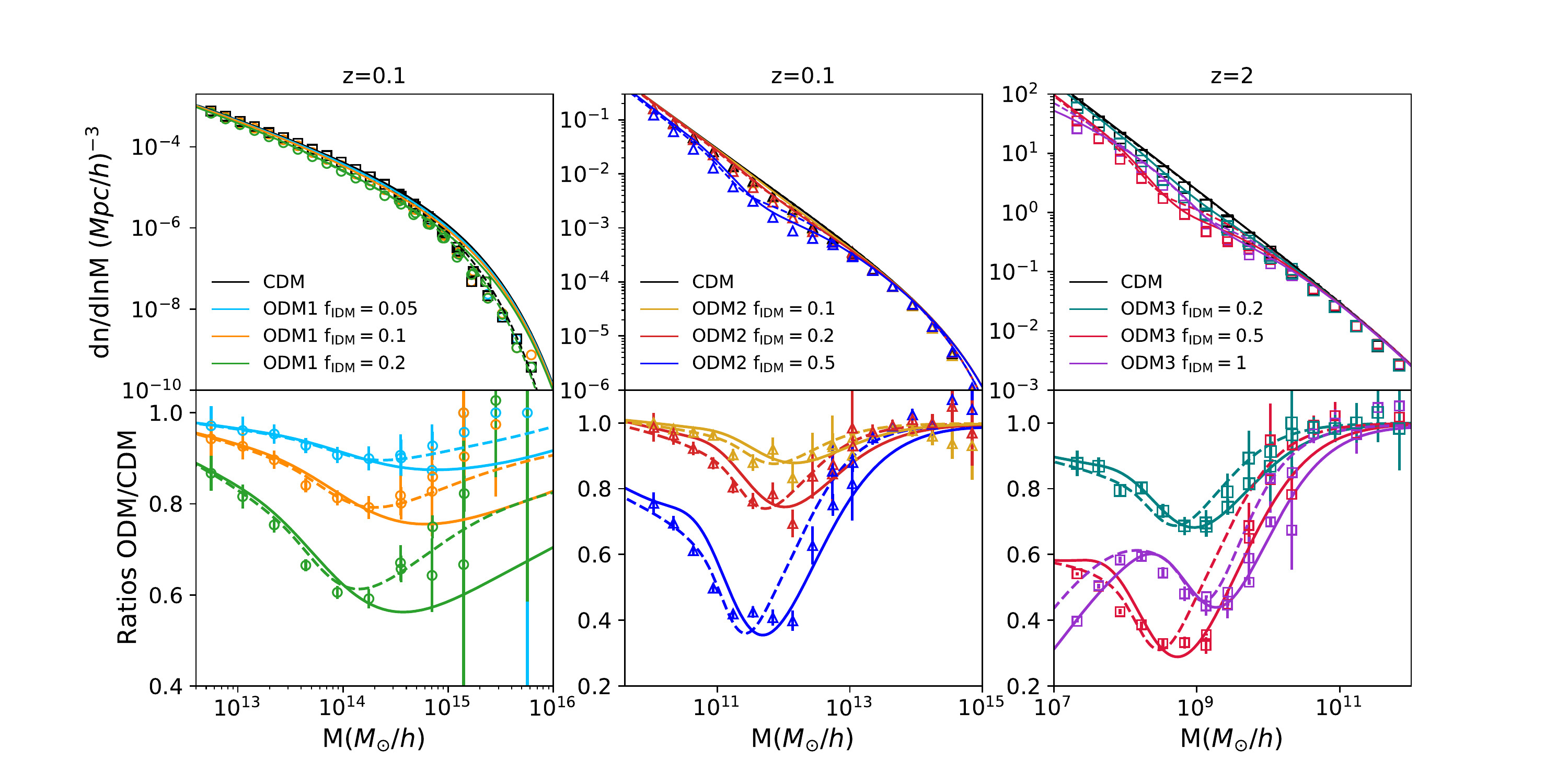}
    \caption{The absolute and relative halo mass functions for all ODM and CDM models (coloured and black). Simulation results are shown as coloured symbols, where the error bars correspond to the relative Poisson errors and all haloes have a minimum of 500 particles. The analytical Sheth-Tormen and smooth-$k$ mass functions are shown as dashed and solid lines.  \textit{Left:} ODM1 at $z=0.1$ with $f_{\rm IDM}=\{0.05, 0.1, 0.2\}$. \textit{Middle:}  ODM2 at $z=0.1$ with $f_{\rm IDM}=\{0.1, 0.2, 0.5\}$. \textit{Right:}  ODM3 at $z=2$ with $f_{\rm IDM}=\{0.2, 0.5, 1\}$.}
    \label{fig:HMFs}
\end{figure*}

\section{Nonlinear Power Spectrum}\label{sec3}
Although the matter power spectrum is not a direct observable, it is generally used as the prime statistic to quantify the effects of cosmic structure formation. 
All power spectra in this paper are obtained with the internal tool of {\tt Pkdgrav3}, which relies on the triangular shaped cloud method and allows for on-the-fly measurements. Note that we checked the convergence of all power spectra from our CDM and ODM simulations by comparing results from different particle resolutions. A summary of the convergence tests is provided in Appendix~ \ref{fig:ConvergenceTestNonLinearP}.

In Fig.~\ref{fig:Non_linear_T-K_ODM} we show the square-root of the ratios between the ODM and CDM power spectra at various redshifts. The nine different panels correspond to ODM1 (left), ODM2 (centre), and ODM3 (right) with increasing fractions of interacting-to-total DM ($f_{\rm IDM}$) from top to bottom. The colour scheme is the same than the one in Fig.~\ref{fig:T-K_ODM}, where the initial transfer functions of the ODM models have been introduced\footnote{Note that the solid line in Fig.~\ref{fig:Non_linear_T-K_ODM} is identical with the transfer functions provided in Fig.~\ref{fig:T-K_ODM}.}. 

Fig.~\ref{fig:Non_linear_T-K_ODM} clearly shows how oscillatory patterns in the power spectrum are washed out and gradually disappear towards lower redshifts. This is a well known trend that has been pointed out before \citep[see e.g.][]{NonLinear_Regime_BAO, NonLinear_Oscillations_Power} and can be attributed to nonlinear mode-coupling. The phenomenon of mode-coupling is best explained in the context of higher-order perturbation theory, where an integral over $k$-modes appears in the equation of motion of the density field in Fourier space. This has the effect of "mixing up" the amplitudes and redistributing them over neighboring k-modes (\citealt{Bernardeau_review}).

The mode-coupling effect described above becomes efficient as soon as modes enter the nonlinear clustering regime. As a consequence, oscillatory features at smaller scales (higher $k$-modes) are smoothed out more efficiently than the ones at larger scales (lower $k$-modes). This trend is reproduced in Fig.~\ref{fig:Non_linear_T-K_ODM}, where the oscillation of ODM1 remain visible until $z\sim 1$, while the ones of ODM2 and ODM3 disappear towards $z\sim3$ and $z\sim 10$, respectively. Below these redshifts, the ODM models do not exhibit any oscillatory features anymore, but they remain distinguishable from CDM due to a smooth suppression of power towards small scales.

\section{Halo Mass function}\label{sec4}
Another important statistic that we investigate in this paper is the halo mass function, i.e. the number density of haloes as a function of their total mass. We thereby define a halo as a density peak with enclosed mass corresponding to 200 times the critical density of the universe ($\rho_{c}$).



\subsection{Simulation results}
We use both the ${\tt AHF}$ (\citealt{AHF}) and the ${\tt Rockstar}$ (\citealt{Rockstar}) codes to find haloes in the outputs of our CDM and ODM simulations. Although these codes rely on different algorithms for the halo-finding process, we have verified that they provide well converged results not only for the CDM but also for the ODM simulations. Note that we do not consider sub-halos in our analysis, and we only keep halos with at least 500 particles to guarantee well converged internal properties.

It is well known that N-body simulations initialized with a smoothed density field below a certain scale (i.e with a cutoff in the power spectrum like in WDM or Axionic DM) suffer from a phenomenon called artificial fragmentation. The initial excess of power at small scales (i.e. shot noise) due to the discrete nature of N-body simulations seeds the formation of small overdense clumps regularly spaced in filaments \citep{WhiteArtifactsHDM}. This induces an upturn in the mass function below the cutoff scale which is not converged when varying the resolution \citep{NonConvergingWDMAurel}. We have looked for potential artifacts in our ODM simulations, using the method described in \citet{LovelWDMhalos} and \cite{AurelSupressedSmallScales}. As a result we did not find any excess compared to the CDM simulations. Note as well that all ODM halo mass functions are well converged between simulations of different resolutions. More details about potential artifacts in ODM, WDM, and CDM simulations can be found in Appendix~\ref{appendixSpurious}.
 
\begin{figure*}
    \centering
    \includegraphics[width =1\textwidth,trim=0.0cm 0.cm 0cm 0.0cm,clip]{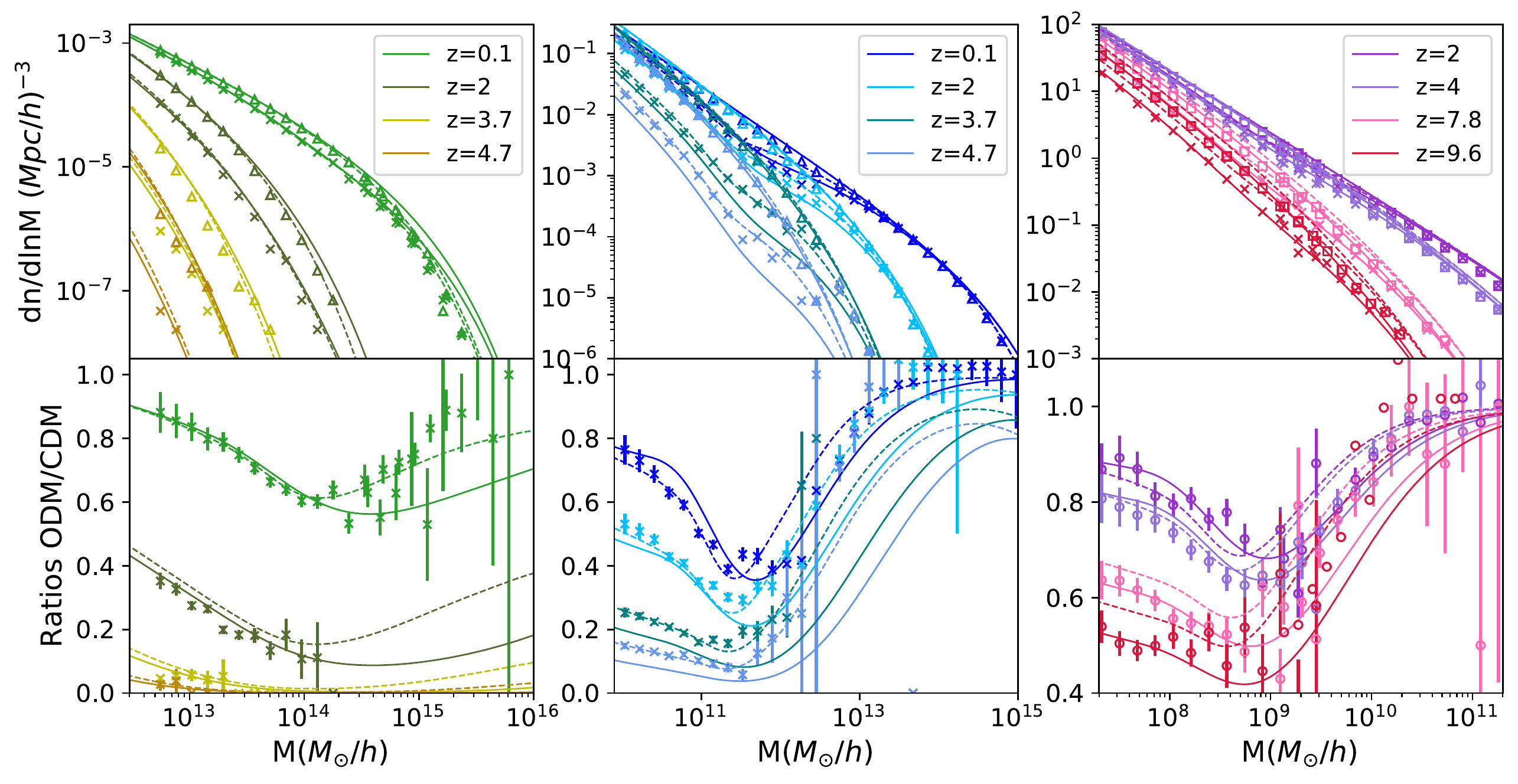}
    \caption{Redshift evolution of the absolute and relative halo mass functions for ODM and CDM models. Symbols with error bars correspond to measurements from simulations. The dashed and solid lines show the Sheth-Tormen and the smooth-$k$ mass functions. For the latter we use the parameters $ (\beta, c)=(3, 3.3)$. From left to right: ODM1 with $f_{\rm IDM} = 0.2$, ODM2 with $f_{\rm IDM} = 0.5$, and ODM3 with $f_{\rm IDM} = 0.2$.}
    \label{fig:HMF_Redshift_Evolution_3panels}
\end{figure*}


The halo mass functions from our CDM and ODM simulations are shown as circle, triangle, and square symbols in Fig.~\ref{fig:HMFs}. The nine different colours refer to all our ODM models assuming the same colour scheme than in Figs.~\ref{fig:T-K_ODM} and \ref{fig:Non_linear_T-K_ODM}. Note that in order to minimise noise due to low-number statistics, each halo mass function is constructed from two simulations with $1024^{3}$ particles and different box lengths (see Table~\ref{tab:SimulationSetup} for a list of our simulations). In the bottom panels of Fig.~\ref{fig:HMFs}, we furthermore plot the ratios between ODM and CDM including Poisson error bars (normalized to the CDM mass function value in each mass bin). ODM1 and ODM2 mass functions are shown at $z= 0.1$ and ODM3 at $z= 2$. 

The most important conclusion that can be obtained from Fig.~\ref{fig:HMFs} is that oscillatory features remain visible in the halo mass function down to low redshifts. This is surprising especially when considering that at the same redshifts oscillations are completely washed out in the power spectrum. When looking at the ratios of the ODM to CDM mass functions (bottom panels), all nine models show clear oscillatory features. This is even true for the ODM1 case with $f_{\rm IDM}=0.05$, where the oscillations in the linear power spectrum are only of the order of a few percent. Considering the absolute halo mass functions, oscillatory features become harder to detect but remain clearly visible at least for the ODM2 and ODM3 models (in agreement with previous works, see e.g. Fig.1 from \citealt{Paranjape}). The situation is different for ODM1 mainly because the  oscillatory feature coincide with the knee in the mass function caused by the exponential cutoff towards very large halo masses.

In Fig. \ref{fig:HMF_Redshift_Evolution_3panels} we plot the redshift evolution of the mass functions for CDM and three of the interacting models (from left to right: ODM1 with $f_{\rm IDM}=0.2 $, ODM2 with $f_{\rm IDM}=0.5 $ and ODM3 with $f_{\rm IDM}=0.2 $). Surprisingly, the oscillating feature is less discernible as moving towards higher redshifts. For ODM1, the oscillatory feature is damped by the exponential tail and what remains visible is an overall depletion of halos. For ODM2 and ODM3 the ratios of the mass functions tend to look like a step function  at high redshifts. This might be due to the decrease of small halos in CDM at late time (for instance looking at Fig. \ref{fig:HMF_Redshift_Evolution_3panels} middle upper panel, the CDM line at $z=0.1$ and $z=2$), such that CDM and ODM abundances get closer to each other, enhancing the oscillation visible in the ratio plots.





This result is interesting: realistic  interactions taking place in the dark sector at early time can leave an imprint on the abundance of virialized structures at late times, adding an oscillating feature on top of the standard power law scaling at small scales. This kind of effect can lead to strong constraints on $f_{\rm IDM}$, but has to be confronted to uncertainties regarding baryonic physics since the halo mass function is not a direct observable.

\subsection{Press-Schechter formalism}
The extended Press-Schechter (EPS) approach provides a convenient semi-analytical tool for predictions of the halo mass function. Over the last decades, it has been shown to not only provide accurate results for $\Lambda$CDM cosmologies (\citealt{PS,Bond_Estathiou_Kaiser_Exc_set,STellipsoidalcollapse1999}) but also for alternative DM models such as warm (\citealt{WDM_Sharpk_Benson,HMF_and_Free-streaming-scale})  and mixed \citep{AurelSupressedSmallScales}, or fuzzy (\citealt{Marsh_Fuzzy_DM,Schneider_21cm}) dark matter. Here we briefly summarize the recent development of EPS models for alternative DM scenarios before discussing their validity regarding ODM models.

In the EPS formalism the halo mass function can be written as
\begin{equation}
  \frac{dn}{d\ln{M}} = -\frac{\bar{\rho}}{2M}f(\nu) \frac{d \ln{\sigma^{2}}}{d\ln{M}} ,
  \label{eq:PS_General_Expression}
\end{equation}
where $f(\nu)$ is the first crossing distribution. For the case of ellipsoidal collapse it is given by
\begin{equation}
f(\nu) = A\left(1+\frac{1}{q\nu^{p}}\right)\sqrt{\frac{2q\nu}{\pi}}\exp{(-q\nu /2 )},
\end{equation}
with $A=0.322$, $p=0.3$, and $q=1$. However, it has been shown that a corrected value of $q=0.707$ provides better match to simulations \citep{STellipsoidalcollapse1999}. The peak height of linear perturbations $\nu=\delta^{2}_{c}/\sigma^{2}(M,z)$ is obtained from the variance 
\begin{equation}
\sigma^{2}(R,z)=\int \frac{k^{2}}{2\pi^{2}} W^{2}(k,R) P_{\rm lin}(k,z) \mathrm{d}k,
\end{equation}
where $W(k,R)$ corresponds to a window function smoothing the density field at a characteristic scale $R$.

There are different options for the window function. The most common and physically motivated filter is the spherical window function in real space (or top-hat filter), which takes the following form in Fourier space:
\begin{equation}
W_{\rm tophat}(k,R) = 3(\sin{kR}-kR\cos{kR})/(kR)^{3}
\end{equation}
and comes with the natural enclosed mass definition $M=4\pi\bar\rho R^3/3$.
Using a top-hat filter leads to reasonable predictions of the halo mass function in $\Lambda$CDM, but overestimates the abundance of small halos in DM models with small-scale power suppression. As shown by \cite{AurelSupressedSmallScales}, this problem can be solved by switching to a sharp-$k$ filter, i.e. a spherical window function in Fourier space. However, the sharp-$k$ mass function fails for models with strong oscillations in the linear power spectrum. Instead of the damped oscillations seen in simulations, it predicts very sharp oscillations of the halo mass function (see the right panel of Fig. \ref{fig:Comparison_Smooth_K_param_Final}). To overcome this problem, \cite{Leo_smooth_k} introduced the smooth-k filter
\begin{equation}
W_{\mathrm{smooth-k}}(k,R)=(1+(kR)^{\beta})^{-1}
\end{equation}
that comes with the mass-to-radius relation $M=4\pi\bar\rho(cR)^3/3$. Both the mass parameter $c$ and the smoothing parameter $\beta$ are free and have to be calibrated to simulations. The $q$-parameter, on the other hand, is left at its original EPS value $q=1$. Note that the smooth-$k$ filter is a generalisation of the sharp-$k$ filter in the sense that the latter can be recovered by formally setting $\beta=\infty$.

In Figs. \ref{fig:HMFs} and \ref{fig:HMF_Redshift_Evolution_3panels} we plot both the Sheth-Tormen (tophat with $q=0.707$) and the smooth-$k$ mass functions as dashed and solid lines, respectively. For the latter we use our best-fitting parameters $(\beta,c)=(3,3.3)$ and keep $q=1$ as mentioned above. We note that the ODM1 models are better matched with a tophat filter, whereas the ODM3 models are more in agreement with the smooth-$k$ filter. For the ODM2 models both filters provide a comparable match to the data. Note, furthermore, that the smooth-$k$ mass function overestimates the abundance of very large halos (see e.g. the left panel of Fig .\ref{fig:HMFs}). This discrepancy can be solved by setting $q=1.2$. However, a higher values for $q$ comes at the cost of a poorer match to the high-redshift data shown in Fig.~\ref{fig:HMF_Redshift_Evolution_3panels}.

At the end of this section, we want to emphasise that there is no unique set of smooth-$k$ parameters $(\beta,c)$ that simultaneously matches models with dark acoustic oscillations (ODM) and models with a cutoff in the power spectrum (such as WDM). This is because $\beta$ controls simultaneously the small-scale slope of the mass function in models with a suppressed power spectrum and the degree of smoothness of the oscillation in ODM models. As a result, ODM and WDM models have to be fitted separately with two different values of $\beta$. We discuss this point in more detail in appendix \ref{appendixPressSchechter}.

\section{Towards constraints from observations}\label{sec5}
We will now investigate to what extent dark acoustic oscillations could leave an imprint on astrophysical observables. We will focus on the cluster mass function, the stellar-to-halo mass relation, and the Lyman-$\alpha$ 1D flux power spectrum. Note that we also investigated other observables such as galaxy and CMB weak lensing as well as the HI velocity function. While ODM models cause a general suppression of power at small angular scales in these measurements, we have checked that any oscillatory features are either washed out or completely dominated by systematics from poorly quantified feedback effects. We therefore do not further discuss these findings here.

\begin{figure}
    \centering
    \includegraphics[width=0.4\textwidth,trim=0.2cm 0.0cm 0.1cm 0.0cm,clip]{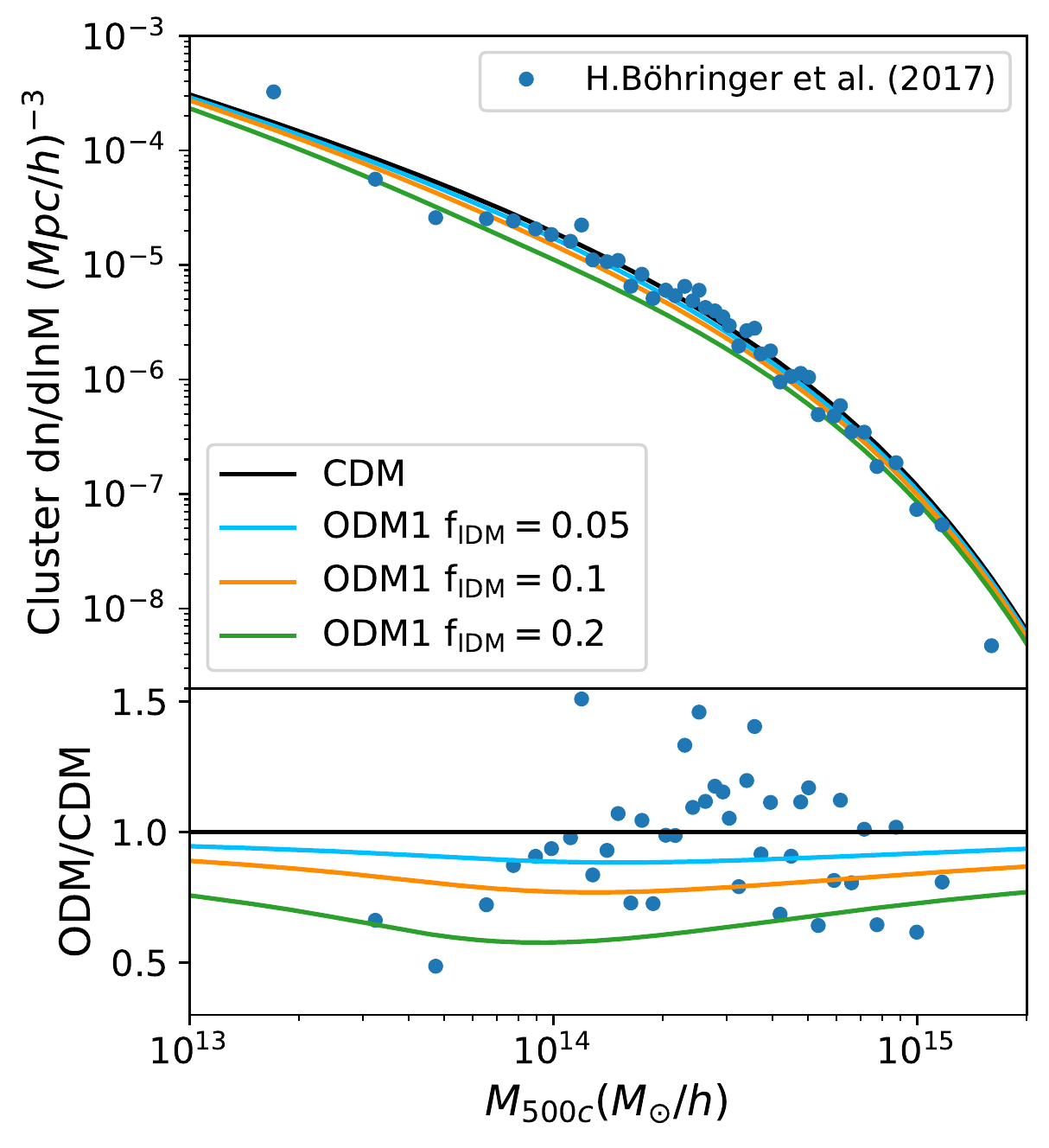}
    \caption{The number density of clusters as a function of their mass $M_{500c}$ estimated from X-ray data. Observational data points (blue circles) are taken from the analysis of \citealt{Rosita_Ref}. Solid lines show the theoretical cluster mass functions for CDM and ODM models, assuming the \citealt{Rosita_Ref} best-fit cosmological parameters. The ratios between the ODM and CDM results are plotted in the lower panel, along with \citealt{Rosita_Ref} data points normalized to the CDM line.}
    \label{fig:ClusterMassFunction}
\end{figure}

\subsection{The cluster mass function}
The mass function of galaxy clusters is a powerful cosmological probe provided the total cluster mass can be accurately determined. One popular way to determine the total halo mass of clusters is via X-ray measurement of gas which can be related to the total halo mass via the assumption of hydrostatic equilibrium \citep[see e.g.][]{Cluster_REview_Pratt2019}. Recent measurements of the X-ray cluster mass function based on data from the REFLEXII galaxy cluster survey can be found in \citealt{Rosita_Ref}.

\begin{figure*}
    \centering
    \includegraphics[width =1\textwidth,trim=0cm 0.0cm 0.0cm 0.0cm,clip]{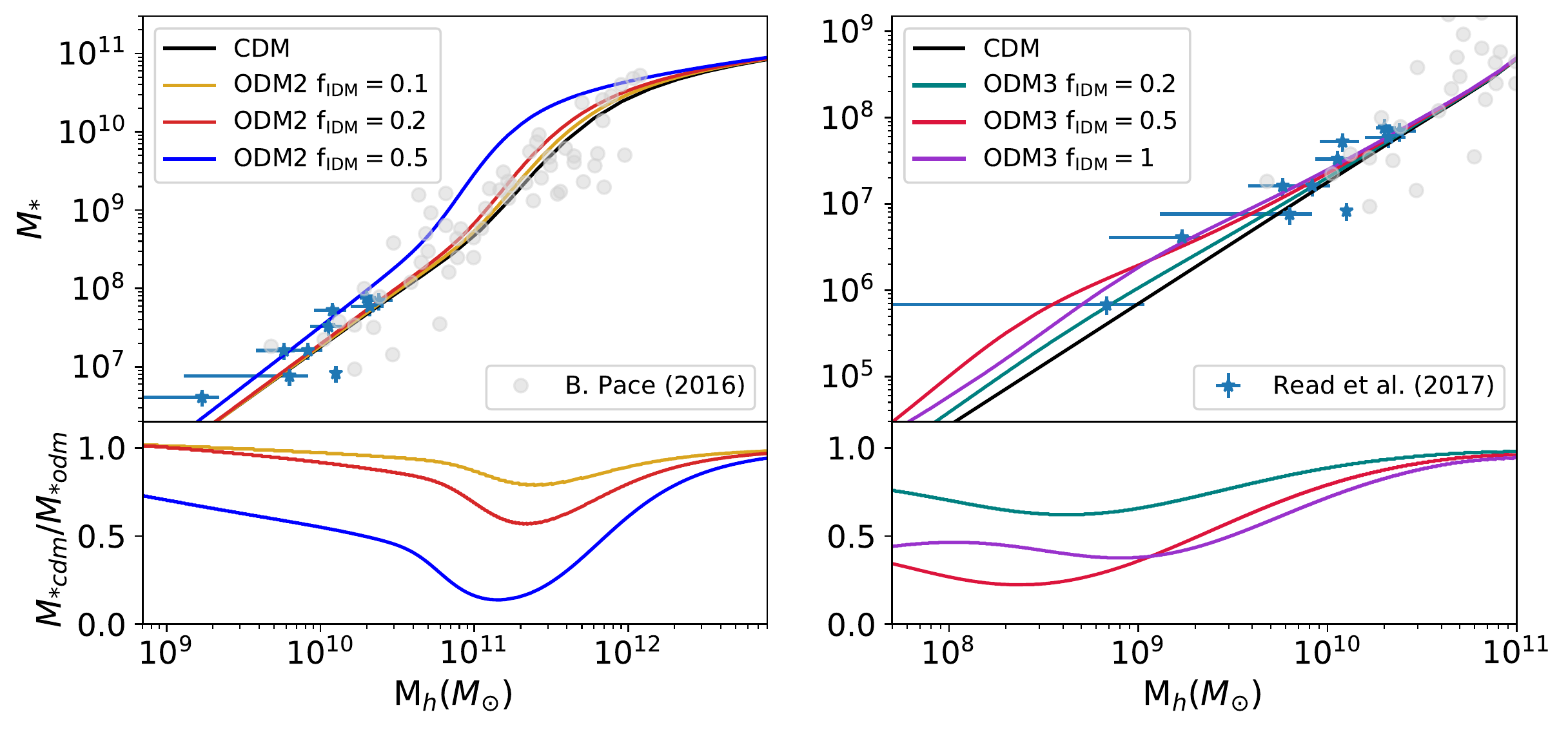}
    \caption{Predicted stellar-to-halo mass relations for CDM (black lines, see \citet{Behroozi_2013}) and ODM (coloured lines, derived as described in section 5.2). The data points correspond to direct measurements from \citet{Andrew_Rotation_Curve} (grey circles) and \citet{Read_et_al_Dwarfs} (blue stars with error bars) based on fitting HI rotation curves from field galaxies. The bottom panel shows ratio between the CDM and ODM stellar-to-halo relations.}
    \label{fig:Abundance_Matching}
\end{figure*}


In order to compare our ODM models to the observed cluster mass function, we take the best-fitting cosmological parameters from REFLEX-II ($\Omega_{m}=0.28$, $\sigma_{8}=0.775$), before calculating the corresponding linear power spectra with CLASS. We then compute the Sheth-Tormen mass functions for both the CDM and ODM1 models at redshift 0.1. The  $M_{500c}$ masses are obtained from $M_{200c}$ assuming a NFW profile (with concentrations from our simulations) and accounting for an additional hydrostatic mass bias of 10 percent. 

In Fig.~\ref{fig:ClusterMassFunction} we plot the cluster halo mass function from REFLEX-II (blue data points) and compare it to the EPS predictions for CDM (black line) and the three ODM1 models (coloured lines). Not surprisingly, the ODM models show a 10-40 percent suppression of the cluster abundance with respect to the CDM case. The ODM oscillations, on the other hand, are very smooth, half an oscillatory mode extending over several orders of magnitude in halo mass. A visual comparison of the ODM cluster mass functions with the data from REFLEX-II indicates that these oscillations are too broad to stand out as a distinctive feature in the data. As a consequence, any ODM signal is likely to show at least some degeneracies with other cosmological or astrophysical parameters.

It would be interesting to quantify the constraining power of cluster mass functions for ODM scenarios by performing a more rigorous MCMC analysis. Especially the upcoming data from the eROSITA all-sky survey \citep{eROSITA} is promising in that respect. However, this exercise goes beyond the scope of this paper and we leave it for for future work.

\subsection{The Stellar-to-Halo mass relation}
The oscillation visible in the halo mass function (Fig. \ref{fig:HMFs}) could also be observable in the Galaxy Stellar Mass Function (GSMF). However, to connect the HMF to the GSMF, we have to assume a function for the star-formation efficiency which can be expressed by the stellar-to-halo mass relation (SHM). This function is only poorly constrained by observations.


Direct measurements of the SHM relation are difficult because they do not only rely on the stellar mass but also on the underlying total halo mass. The latter can be estimated using for example information from X-ray emission, from gravitational lensing, or from rotation-curve measurements (\citealt{Star_to_Halo_mass_Mandelbaum,Star_to_Halo_mass_Gavazzi_2007,H13_WL_Star_Halo_Rel,Star_to_Halo_mass_Uitert,Read_et_al_Dwarfs}). Although these SHM measurements are subject to many potential systematics, they show an encouraging agreement with results from abundance matching the observed luminosity functions to the simulated CDM halo mass functions (\citealt{Behroozi_2010,Behroozi_2013,Star_to_Halo_Careful_High_mass_Kravtsov2018}).

A rigorous abundance matching procedure assuming a $\Lambda$CDM mass function and including halo-to-halo scatter has been performed by \cite{Behroozi_2013}. The five-parameters fit of the resulting SHM relation is shown as black line in Fig. \ref{fig:Abundance_Matching}. To predict the shape of the SHM in the ODM models, we match the CDM and ODM cumulative mass functions  (using the smooth-$k$ parametrisation from section \ref{sec4}). This yields a relation between $M_{\rm cdm}$ and $M_{\rm odm}$. We then convolve the SHM from \cite{Behroozi_2013} with this relation in order to obtain the correct stellar-to-halo relation for ODM models.

In Fig.~\ref{fig:Abundance_Matching} we plot the relation between the stellar and halo mass for the ODM2 (left) and ODM3 models (right). In both cases we see a broad oscillatory feature with a minimum at $10^{11}-10^{12}$ M$_{\odot}$ for ODM2 and $10^8-10^9$ M$_{\odot}$ for ODM3. Note that the results for ODM1 are not shown here because they remain very close to CDM with differences far below the current observational uncertainties.

Next to the ODM and and CDM abundance matching results of Fig.~\ref{fig:Abundance_Matching}, we show direct measurements of the stellar-to-halo mass relation from  \cite{Andrew_Rotation_Curve} (grey dots) and \cite{Read_et_al_Dwarfs} (blue dots with error-bars). Both measurements are based on the analysis of HI rotation curves from individual field galaxies, where the total halo mass is estimated based on the DC14 \citep{DC14} and the coreNFW \citep{CoreNFW} halo profiles. Although both these profiles are well tested with hydrodynamical simulations, it is worth noting that the results may be subject to hidden systematics (see e.g. \citealt{Sebastian_2018} for a more detailed discussion). Regarding the measurement from \cite{Andrew_Rotation_Curve}, we only plot the \emph{well-constrained} sub-sample from the paper. While this sample only contains galaxies with well behaved posteriors regarding the halo mass estimates, one should keep in mind that such an \emph{a posterio} selection could lead to a bias with respect to the mean stellar-to-halo-mass relation.

Given the potential systematics of the direct measurements and the cosmology dependence of the abundance matching technique, it is difficult to draw final conclusions regarding the underlying DM scenarios. Most of the ODM and the CDM model are in good agreement with the measurements from \cite{Andrew_Rotation_Curve}. One exception is the ODM2 model with $f_{\rm IDM}=0.5$ that shows considerable tension with the galaxies residing in about $\sim 10^{11}$ M$_{\odot}$ haloes. The measurements from \cite{Read_et_al_Dwarfs} seem to be slightly better matched by the ODM3 models compared to CDM. Note, however, that these results are based on an extrapolation of the abundance-matching line from \cite{Behroozi_2013} which is highly uncertain.

While current stellar-mass to halo-mass measurements based on HI rotation-curves might suffer from considerable systematics, future radio interferometers will provide a wealth of new data with resolved rotation curves from thousands of blindly selected galaxies (see e.g. the discussion in \citealt{Sebastian_Aurel_Cosmo_constraints}). This data will help to strongly constrain the SHM relation and can be used to search for potential signatures of ODM models.

\begin{figure*}
    \centering
    \includegraphics[width =1.\textwidth,trim=0.0cm 0.0cm 0.0cm 0.0cm,clip]{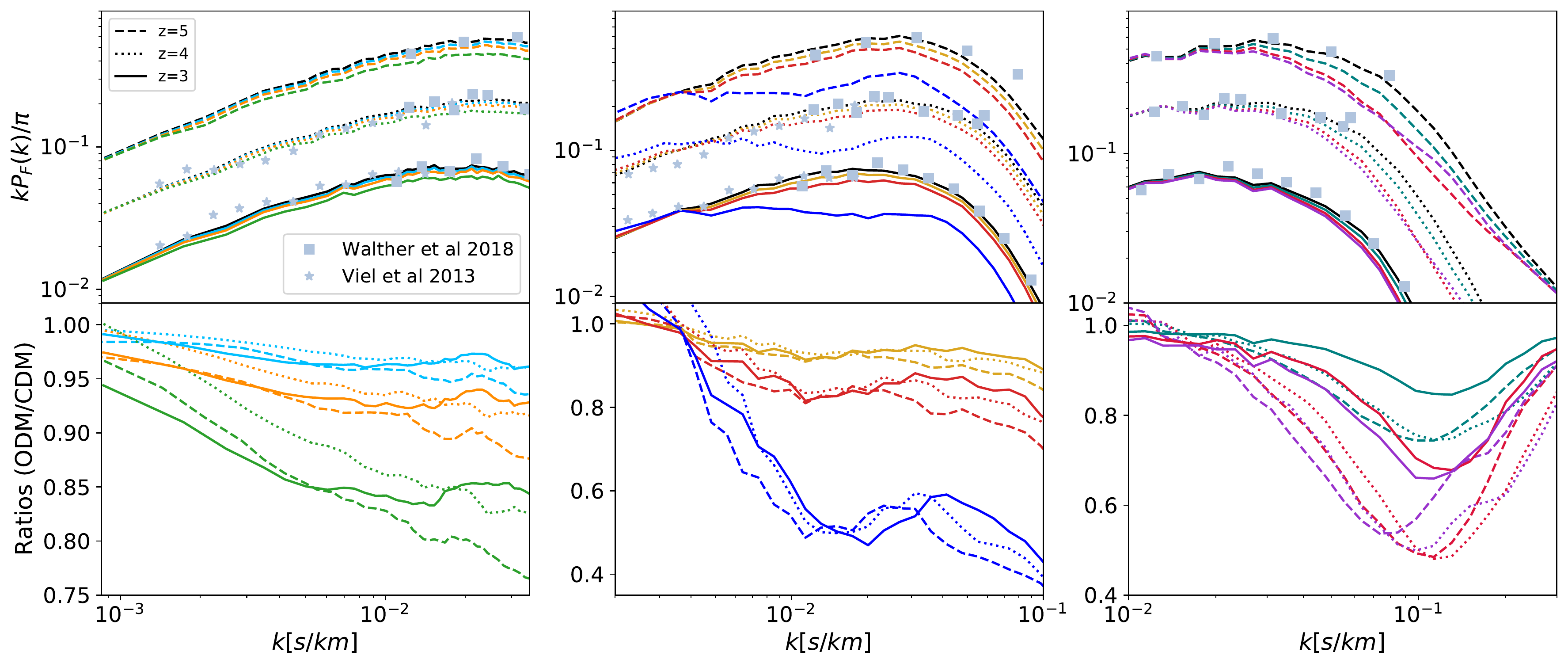}
    \caption{Lyman alpha 1D flux power spectra for CDM (black lines) and ODM models. Each color corresponds to a different fraction of interacting-to-total DM (same color scheme as in Fig .\ref{fig:T-K_ODM}). \textit{Left:} ODM1 with $f_{\rm IDM}=\{0.05, 0.1, 0.2\}$. \textit{Middle:}  ODM2 with $f_{\rm IDM}=\{0.1, 0.2, 0.5\}$. \textit{Right:}  ODM3 with $f_{\rm IDM}=\{0.2, 0.5, 1\}$. The different redshift bins with $z=3$, 4, and 5 are shown as dashed, dotted, and solid lines. The observational data in grey is obtained from \citet{Walther_2019} (squares) and \citet{Lyman_Alpha_Viel} (stars). }
    \label{fig:LymanAlpha}
\end{figure*}

\subsection{Lyman-\texorpdfstring{$\alpha$}{Lg} forest}
One of the most powerful observables to constrain non-CDM dark matter models is the Lyman-$\alpha$ 1-D flux power spectrum $P_{F}(k,z)$ \citep[e.g.][]{Seljak_2006_LyA_SterileNu, Viel_2006_LyA_SterileNu, Lyman_alpha_Boyarsky_2009, Lyman_Alpha_Viel, Lyman_AlphaBaur_2017, Lyman_alpha_2017}. In this section we will investigate to what extent $P_{F}(k,z)$ can be used to detect dark acoustic oscillations. Previous studies have looked into the effects of interacting DM-DR models on the Lyman-$\alpha$ forest \citep{DM-nu_Boehm,Lymanalpha_ethos}, but they were limited to scenarios with oscillatory features at very small scales. Such models are often similar to the well-studied case of warm DM, where the Lyman-$\alpha$ forest has shown to be a powerful tool leading to strong constraints. Here we will also focus on less pronounced oscillatory features appearing at larger scales, which do not exhibit a strong cutoff in the linear power spectrum and are conceptually very different than the standard warm DM scenario.

We use $N$-body simulations run with {\tt Pkdgrav3} to predict the Lyman-$\alpha$ flux power spectrum $P_{F}(k,z)$. The simulations are characterised by $L_{\mathrm{box}}=64$ Mpc/h for ODM1 and ODM2, and $L_{\mathrm{box}}=20$ Mpc/h for ODM3, all with particle numbers $N=512^3$. This corresponds to a simulation particle mass of $m_{\rm p}=1.7 \cdot10^{8}$ $\rm M_{\odot}/\rm h$ and $m_{\rm p}=2.2 \cdot10^{7}$ $\rm M_{\odot}/\rm h$. Next to all the nine ODM models introduced above, we also run two CDM simulations with the same setup.

In order to generate smoothed, one-dimensional density and velocity fields, we randomly select line-of sight skewers parallel to the simulations box axes. We bin the skewers and interpolate the DM particle properties by assigning Gaussian weights
\begin{equation}
w_{p,v} = \exp{\left[-r^{2}/(2\sigma^{2}_{p,v})\right]}
\end{equation}
to every DM particle, where $r$ refers to the distance of a DM particle to a given grid-point on the skewer. The standard deviations $\sigma_{p}$ and $\sigma_{v}$ designate the particle and velocity pressure smoothing scales. We obtain good agreement with observational data for $\sigma_{\rm p} = 58$ ckpc. For the velocity field, we use $\sigma_{\rm v} =228  $ ckpc, as suggested by \citealt[]{Lyman_A_Collisionless} (see their appendix A).

The local optical depth of the IGM is modelled using the fluctuating Gunn-Peterson approximation 
\begin{equation}
    T\sim T_{0}\left(\frac{\rho_{b}}{\bar{\rho}_{b}}\right)^{\gamma-1},
    \hspace{1cm}
    \tau_{\mathrm{local}} = A(z)(1+\delta_{b})^{2.7-0.7\gamma},
    \label{eq:T_and_tau}
\end{equation}
where $A(z)$ can be found in \citealt{LyA_Croft_2002} (see their Eq.~2) and depends on the temperature normalisation factor $T_{0}$. The thermal broadening effect is included via the relation 
\begin{equation}
    \tau_{\mathrm{obs}}(u) = A_{\rm norm}\int du' \frac{\tau_{\mathrm{local}}(u')}{b(u')\sqrt{\pi}} \exp{\left(-\frac{(u-u_{0}(u'))^{2}}{b^{2}(u')}\right)},
    \label{eq:Voigt_convolv_tau}
\end{equation}
where $A_{\rm norm}$ is a normalisation factor. The line-of-sight velocity of the gas $u_{0}(u') = u'+u_{\mathrm{pec}}(u')$ is obtained by adding the Hubble flow velocity ($u'$) to the one-dimensional velocity field from our simulations ($u_{\rm pec}$). The thermal broadening width is given by $b(u') =\sqrt{2k_{\rm B}T(u')/m_{\rm p}}$, with $m_{\rm p}$ being the proton mass. For more details about the procedure, see \citet{LyA_Lukic_Reference}. 

The 1-D flux power spectrum $P_{\rm F}(k,z)$ is defined as the two point correlation function in Fourier space of: 
\begin{equation}
    \delta_{F}=\frac{F-\big<F\big>}{\big<F\big>},
    \hspace{0.5cm}
    {\rm with}\hspace{0.5cm}
    F=\exp{\left(-\tau_{\rm obs}\right)}.
    \label{eq:delta_1d_flux}
\end{equation}
We calculate $\delta_{F}$ using  1000 randomly selected skewers from each simulations at redshift 5, 4 and 3. This number is large enough to guarantee converged results for the flux power spectrum. 

As a reference, we take the IGM parameters $T_{0}$ and $\gamma$ from \cite[see their table~4]{Walther_2019}. The factor $A_{\rm norm}$ is normalised to observations and has a slightly different value for each selected skewer. This step is necessary to correct for resolution issues in our large simulation boxes.

The resulting 1D flux power spectra are shown in Fig. \ref{fig:LymanAlpha}. The black curves correspond to CDM, and the coloured ones to the ODM models. The dashed, dotted and solid lines show results from the redshifts 5, 4 and 3. The observational data points are taken from \cite{Walther_2019} (square symbols) and from \cite{Lyman_Alpha_Viel} (star symbols), respectively. While we plot the absolute 1D flux power spectra in the top panels, the bottom panels show the ratios between the ODM and CDM models.

The left-hand panel of Fig. \ref{fig:LymanAlpha} is focused on the ODM1 models. The dominating effect that we see is an overall suppression of power for the ODM1 models with respect to CDM. While no oscillations are visible in the absolute power spectra, a tentative feature can be distinguished in the ratio plot. However, since the oscillation (with peak at $k\sim 2-3\times 10^{-2}$ s/km) has a small amplitude and is only present in the $z=3$ results, it is unclear whether it is really caused by dark acoustic oscillations or by other nonlinear effects. 

In the central panel of Fig.~\ref{fig:LymanAlpha} we show results for the ODM2 models. Next to the general suppression of power in ODM, an oscillatory feature is clearly visible in both the absolute Lyman-$\alpha$ 1D flux (top panel) and in the ratio plot (bottom panel) for the $f_{\rm IDM}=0.5$ model. The other models also show an oscillatory feature, which, however, is of smaller amplitude and therefore only visible in the ratio plot. The  peak of the oscillations signal is visible at all redshifts and is positioned at $k\sim0.04$ s/km. This scale agrees with the first and main oscillation peak in the transfer function, which is at $k\sim 3-5$ h/Mpc (see Fig .\ref{fig:T-K_ODM}). We therefore conclude that the oscillatory features visible in the ODM2 models are indeed a sign of dark acoustic oscillations.
Note that, compared to the results from the 3-D matter power spectrum (see Fig. ~\ref{fig:Non_linear_T-K_ODM}), the oscillations survive until longer, being well visible at $z=3$. This could be due to the fact that the Lyman-$\alpha$ forest flux spectrum preferentially probes low-density regions that become nonlinear at later times. Indeed, highly non-linear regions, located in halos, completely absorb the Lyman-$\alpha$ radiation, and do therefore not contribute to the flux signal.

In the right-hand panel of Fig.~\ref{fig:LymanAlpha}, we plot the results from ODM3. While we again observe a net suppression of power with respect to CDM, clear oscillatory features are neither visible in the absolute power spectrum nor in the ratio plot. Note that at first sight, these results do not seem to fully agree with the findings from \cite{Lymanalpha_ethos}, who investigated the Lyman-$\alpha$ flux power spectrum for a model similar to ODM3 and reported a small oscillatory feature at $k\sim 0.1-0.2$. While we cannot reliably confirm this feature, we do see a slight bump in the ODM3 case with $f_{\rm IDM}=1$ at the same scale, best visible at $z=4$ (purple dotted line). Whether this small bump is a remnant dark acoustic oscillation cannot be fully verified or excluded with our current method and simulation setup.

In summary, we conclude that dark acoustic oscillations can, in principle, be detected with data from the Lyman-$\alpha$ forest. At the same time, all ODM models investigated here lead to a significant suppression of power, especially at smaller scales (above $k\sim 10^{-2}$ s/km). Only a proper MCMC analysis with varying cosmological and astrophysical parameters will be able to tell to what extend ODM models can be excluded by observations and what role oscillatory features might play.


\section{Conclusions}\label{sec6}
In this paper, we investigate how dark acoustic oscillations in the initial transfer function affect prime statistics of nonlinear structure formation, such as the matter power spectrum and the halo mass function. Dark acoustic oscillations are a generic outcome of particle interactions within the dark matter sector \citep{Buckeley_DM_DR,Dark_atom_model,Ethos_eff, AssymetricDM} or between dark matter and the standard model \citep{DM_photons_Boehm,DM_neutrino_int_frac, DM-nu_Boehm}. While interactions involving one dominating dark matter species lead to strong (and often strongly damped) oscillations at very small scales, a more complex DM sector with interactions between sub-components may result in small oscillatory features appearing at different scales in the initial power spectrum.

We use the ETHOS framework \citep{Ethos_eff} to define a set of different oscillatory dark matter (ODM) models, where either all or only a fraction of DM is assumed to interact with a massless dark radiation mediator particle. In this way, we are able to construct three generic scenarios (ODM1, ODM2, and ODM3), where oscillations appear at large, medium, and small physical scales of the initial transfer function. For each scenario, we furthermore assume three different fractions of interacting-to-total DM, ending up with nine benchmark models that make it possible to study  a large variety of oscillatory features at different scales and with different amplitudes.

In order to investigate the evolution of dark acoustic oscillations during the regime of nonlinear structure formation, we run a suite of gravity-only N-body simulations with the code {\tt Pkdgrav3} \citep{stadel2001, pkdgrav3}. We use these simulations to study the nonlinear power spectrum, finding that oscillatory features are gradually washed-out once they enter the nonlinear regime. This effect can be attributed to mode-coupling and has been observed in other contexts before \citep{NonLinear_Regime_BAO, NonLinear_Oscillations_Power}. Note that the oscillations do not disappear instantaneously from the power spectrum. The smoothing happens gradually, leaving room for the possibility that oscillations remain detectable in observations related to the high-redshift power spectrum.

In a next step, we study the effects of initial oscillations on the halo mass function at both high and low redshifts. Surprisingly, we find that oscillatory features are imprinted in the mass function and remain visible even in the strongly nonlinear regime at low redshifts. This is not only true for strong initial oscillations but also for models with subdominant interacting-to-total DM fractions, where the initial oscillations are of much lower amplitude.

Based on our results on the power spectrum and the mass function, we investigate potential observables where dark acoustic oscillations could stand out as a distinctive feature. We first look into the cluster mass function and the stellar-to-halo mass relation, showing that oscillations remain visible but are extended over several orders of magnitude in halo mass. This means that while ODM models can be constrained with both the cluster mass function and with direct measurements of the stellar-to-halo mass relation, they are unlikely to provide a signal that is distinctive enough to avoid degeneracies with other astrophysical or cosmological parameters.

Finally, we also investigate the potential of the Lyman-$\alpha$ forest as a probe for dark acoustic oscillations. We find that oscillations remain visible in the 1D flux power spectrum even for models with rather small initial oscillations. Surprisingly, the oscillatory features seem to survive longer in the Lyman-$\alpha$ flux power spectrum compared to the 3D matter power spectrum. We speculate that this could be due to the fact that the Lyman-$\alpha$ forest is a probe of the low-density environment, where nonlinear structure formation is delayed. We conclude that the Lyman-$\alpha$ flux power spectrum is a promising observable to detect potential dark acoustic oscillations.


\section*{Acknowledgements}
We thank Dough Potter for precious help with Pkdgrav3, Julian Adamek for assistance with Rockstar, Romain Teyssier and Joachim Stadel for useful discussions.  We furthermore thank Maria Archidiacono and Francis-Yan Cyr-Racine for their quick answers and useful comments regarding the ETHOS model and its implementations in the Boltzmann solvers {\tt Camb} and {\tt Class}. This study is supported by the Swiss National Science Foundation via the grant {\tt PCEFP2\_181157}.

\section*{Data Availability}
The data underlying this article will be shared on reasonable request to the corresponding author.





\bibliographystyle{mnras}
\bibliography{example} 




\appendix

\section{Convergence test for the non linear power spectrum}\label{appendixNonlinearPConvergence}
Here, we discuss the convergence of our N-body simulations. We compare the non-linear power spectra from simulations of the same ODM model but with different resolutions, measured with ${\tt Pkdgrav3}$.

\begin{figure}
    \centering
    \includegraphics[width=0.4\textwidth]{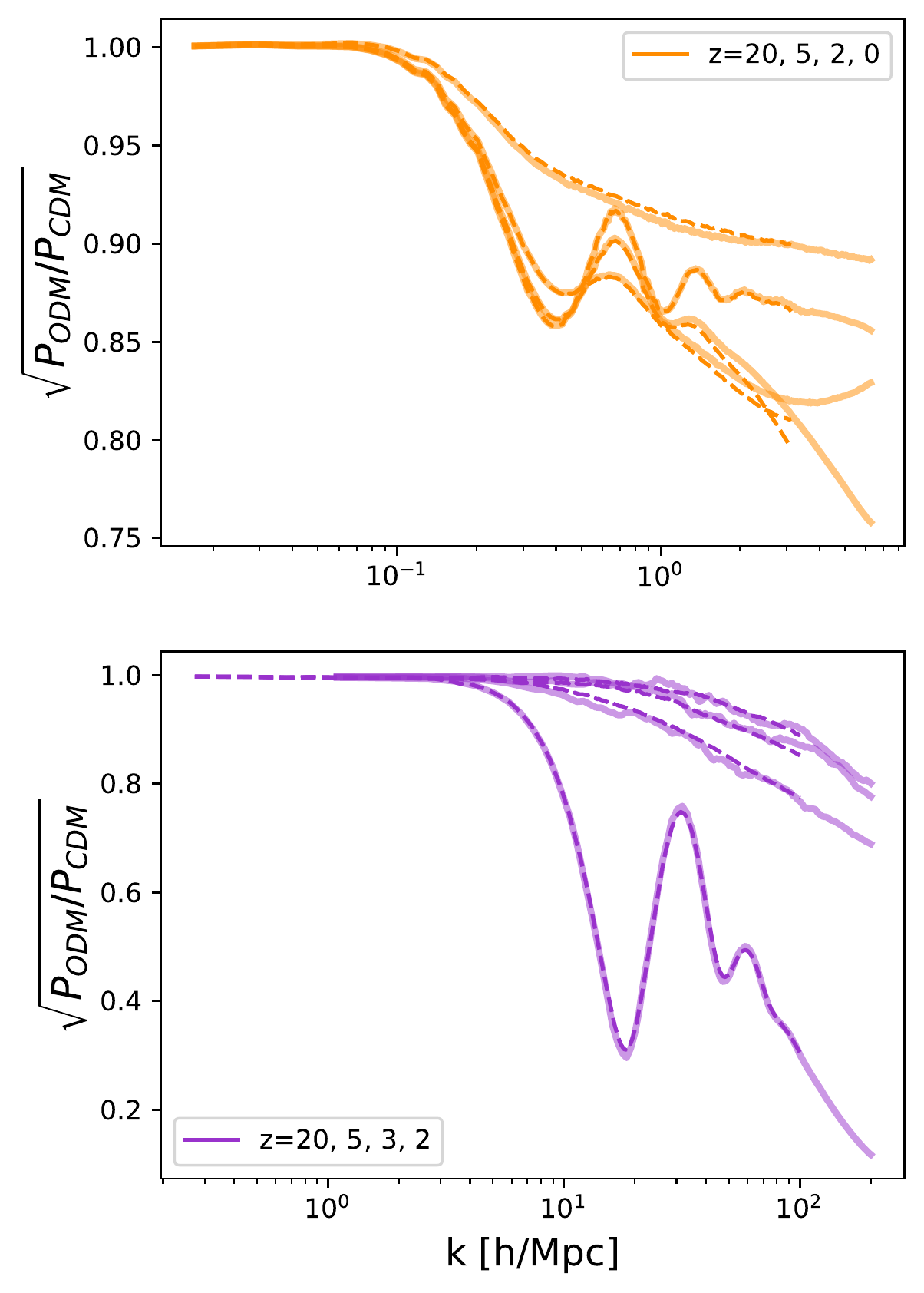}
    \caption{Ratios of the ODM and CDM power spectra for simulations with different resolutions. \textit{Upper panel}: ODM1 with $f_{\rm IDM} =0.1$. The dashed and solid lines are measured from simulations with $L_{\rm box}=512$ Mpc/h and $N_{\rm p}=512^{3}$ and $N_{\rm p}=1024^{3}$, respectively. \textit{Lower panel}: ODM3 with $f_{\rm IDM} =1$. The dashed lines are measured from the runs with $L_{\rm box}=32$ Mpc/h and $N_{\rm p}=1024^{3}$, and the solid line from the runs with $L_{\rm box}=8$ Mpc/h and $N_{\rm p}=512^{3}$.}
    \label{fig:ConvergenceTestNonLinearP}
\end{figure}

In the upper panel of Fig. \ref{fig:ConvergenceTestNonLinearP} we plot the square-root of the ratios between the ODM1 (with $f_{\rm IDM} = 0.1$) and CDM power spectra at various redshifts. The dashed lines are measured from  our simulation with $512^{3}$ DM particles and $L_{\rm box} = 512$ Mpc/h, and the solid lines are from the higher resolution run with the same box size but with $1024^{3}$ DM particles. The power spectra are converged at $z = $ 20 and 0, but slightly differ for the largest k-modes ($k\sim 1-3$ h/Mpc) at $z=$ 5 and 2. These scales are indeed entering the non-linear regime at these redshifts, and the higher resolution simulation can fully capture the effects of mode coupling that distributes power from small to large scales, unlike the lower resolution run. We note the good convergence at every redshift down to the scale of the first peak of the oscillations.

In the lower panel of Fig. \ref{fig:ConvergenceTestNonLinearP} we show the square-root of the ratios between the ODM3 (with $f_{\rm IDM} = 1$) and CDM power spectra. The dashed lines are measured from the simulation with $512^{3}$ DM particles  and $L_{\rm box} = 8 $ Mpc/h and the solid lines from the run with $1024^{3}$ particles and $L_{\rm box} = 32 $ Mpc/h. Note that,  the simulation with the larger number of DM particles has a smaller particle mass, due to its larger box size. The spectra are well converged at these scales and redshifts.

\section{Convergence test for the mass function and analysis of halo properties}\label{appendixSpurious}

\begin{figure*}
    \centering
    \includegraphics[width =1\textwidth]{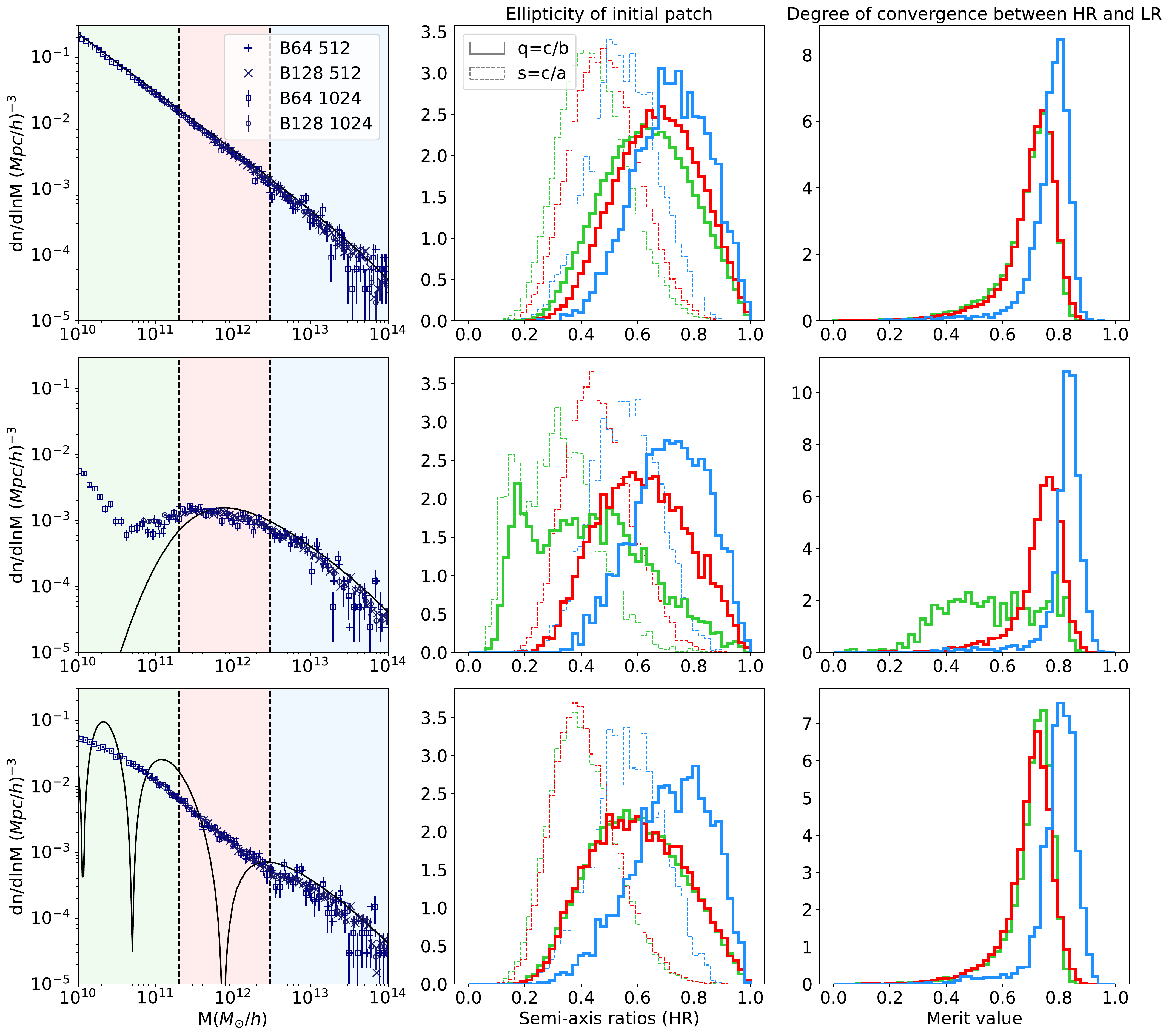}
    \caption{ \textit{Left}: Halo mass function at $z=0$ measured from simulations for CDM, WDM, and ODM2 with $f_{\rm IDM}=1$. Plus and cross signs come from LR catalogs ($N_{\rm p}=512^{3}$) with $L_{\rm box}=64$Mpc/h and $L_{\rm box}=128$Mpc/h, respectively. Square and circles signs are extracted from HR halo catalogs with same box sizes but with $N_{\rm p}=1024^{3}$. The black curve shows the Press-Schechter sharp-$k$ mass function with c=2.5. The three coloured regions indicate the three halo mass bins from which we measure the halo properties shown in the middle and right-hand columns. \textit{Middle}: Distribution of ellipticities (expressed as semi-axis ratios q and s) of the Lagrangian regions in the initial conditions that later collapse to form halos at $z=0$ in the two HR catalogs. The three colors correspond to the mass bins shown as coloured regions in the left-hand panels. The presence of numerical artifacts is manifest in WDM but not in ODM. \textit{Right}: Degree of convergence of halos in the LR simulation, compared to the HR simulation, as explained in the text (see \ref{appendixSpurious}). The distributions are again binned according to the coloured regions shown in the left-hand panels. The merit function is computed for halos in the LR simulations: for each individual halo we look for the equivalent halo in the HR simulation, and compute the fraction of volume shared between the two halos. A small merit means that a halo is not converged, and is likely to be spurious. This reveals the presence of spurious halos in WDM, but not in ODM.}
    \label{fig:Halo_Analysis_Appendix_Spurious}
\end{figure*}

In this section, we look into the properties of ODM halos to make sure that they are well converged and that ODM simulations are not polluted by numerical artifacts. Our analysis is based on the work of \citet{LovelWDMhalos} and \citet{AurelSupressedSmallScales}.

We focus on three models: CDM, ODM 2 (with $f_{\rm IDM}=1$) and WDM with a thermal mass of 0.25keV. We run low resolution (LR, with $N_{\rm p} = 512^{3}$) and high resolution (HR, with $N_{\rm p} =1024^{3}$) simulations with $L_{\rm box}= 128$ Mpc/h and $L_{\rm box}= 64$ Mpc/h. We measure halos with {\tt AHF} \citep{AHF}. We exclude sub-halos, and  only consider halos with at least 300 particles. This makes a total of 12 different halo catalogs (4 per model). We trace the particles inside each halo back to the initial conditions, and look at the 3D geometry of the patch they form. Furthermore, we perform a convergence test between the LR and HR halo catalogs.

To measure the ellipticities of Lagrangian volumes that later collapse to form halos we repeat the procedure of \cite{AurelSupressedSmallScales}. We trace the particles inside each halo back to the initial conditions and measure the semi-axis ($a,b,c$ with $a\ge b \ge c$) of the 3D patch they form, using the inertia tensor. We then define
the semi axis ratios $s=c/a$ and $q=c/b $. Spherical volumes correspond to values of $s$ and $q$ close to 1, prolate or oblate regions to values of $s$ and $q$ close to 0. Here, we only use halos from the HR catalogs.

For the convergence test, we find for each halo $H_{1}$ from a LR simulation the equivalent halo $H_{2}$ in the HR simulation with the same box size (see e.g. \citet{AurelSupressedSmallScales}). We then define a merit function that describes "how converged" a halo is: $M_{c}=V_\textsubscript{shared}/{\sqrt{V_{1}V_{2}}}$, with $V_{\rm shared}$ the volume shared between the two halos and $V_{1}$ and $V_{2}$ the volumes of the LR and HR halos, respectively. We expect non-converged numerical artifacts to have small values of $M_{c}$.

In the left-hand column of Fig. \ref{fig:Halo_Analysis_Appendix_Spurious}, we show the mass functions measured from the simulations for CDM, WDM and ODM along with the corresponding sharp-k mass functions. We systematically overplot data from the LR (plus and cross signs) and HR (square and circles signs with Poisson error bars) halo catalogs.

In the middle column of Fig. \ref{fig:Halo_Analysis_Appendix_Spurious}, we plot the binned ellipticity distributions of the Lagrangian regions that later collapse to form halos in the two HR catalogs. The colors (green, red, blue) correspond to the halo mass bins, from which each distribution is measured. The mass binned are shown as coloured regions in the left-hand columns. We stress on several things:

\begin{itemize}
\item The bigger the halo, the more spherical the initial patch. This is expected: Heavier halos originate from larger regions which have therefore semi-axis ratios closer to 1.
\item ODM halos around the first trough of the oscillation (red region of Fig. \ref{fig:Halo_Analysis_Appendix_Spurious}) originate from slightly more ellipsoidal patches than where CDM originates. A suppression of power at a given scale $R$ has indeed an effect on the sphericity of overdense regions in the initial conditions with volume $V(R)=4\pi R^{3}/3$ that later collapse to form halos with mass $M=\bar{\rho}_{m}V(R)$. However, this effect is small. 
\item  WDM exhibits a bi-modal distribution as already noted by \cite{AurelSupressedSmallScales} and \citealt{LovelWDMhalos}. Halos below the spurious upturn ($M<10^{11} M_{\odot}/h$ where artificial fragmentation dominates) arise from very elongated patches, with a peak close to ($s,q=0.15$). We note that halos around the peak of the sharp-k mass function ($10^{11}< M<10^{12} M_{\odot}/h$) arise  from  ellipsoids with reasonable properties.
\end{itemize}

\begin{figure*}
    \centering
    \includegraphics[width =1\textwidth]{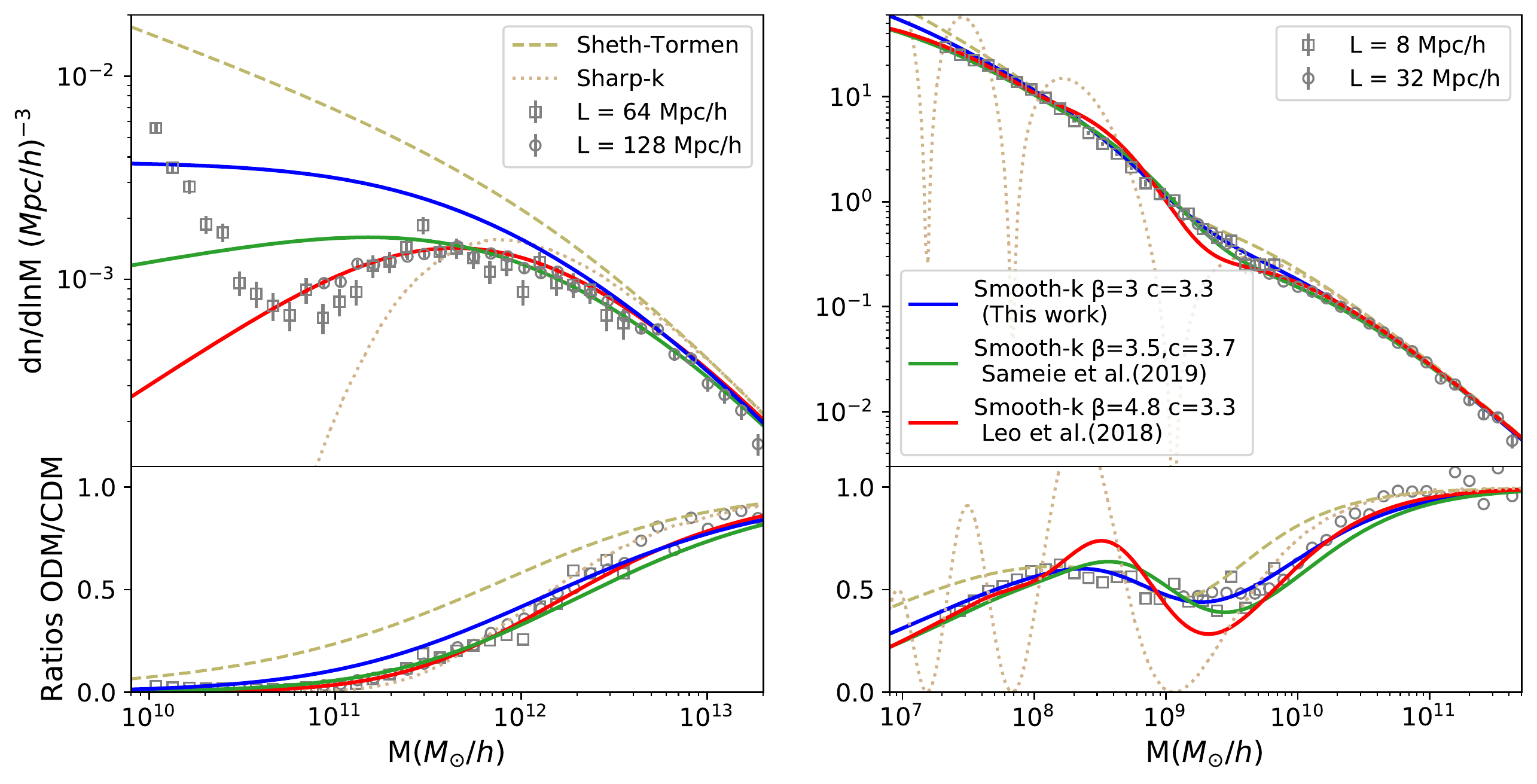} 
    \caption{Halo mass functions for WDM ($m_{\rm th}=0.25$ keV) and ODM3 ($f_{\rm IDM}=1$). The tophat (Sheth-Tormen) and sharp-$k$ mass functions are shown as dashed and dotted lines. The coloured solid lines correspond to the smooth-$k$ mass function with three different ($c,\,\beta$) parameter values. The simulation results are shown as grey data points with Poisson error bars. For the smooth-$k$ filter, there is no unique set of ($c,\,\beta$) parameters that is able to simultaneously fit both the WDM and ODM3 models.}
    \label{fig:Comparison_Smooth_K_param_Final}
\end{figure*}

In the right-hand column of Fig. \ref{fig:Halo_Analysis_Appendix_Spurious}, we plot the distributions of merit function values for the halos in the three mass bins. We note that ODM halos have a distribution similar to CDM. WDM halos below the upturn have low values of the merit function, as expected. 

We conclude that ODM simulations do not produce a significant number of numerical artifacts. They are converged and originate from reasonable ellipsoids in the initial conditions. We can therefore trust the halo mass functions measured in our simulations.

\section{Comparison of our Press-Schechter parametrization with other works}\label{appendixPressSchechter}

In this section, we will develop a statement made in section \ref{sec4}, that one cannot find a unique set of smooth-k parameters to fit at the same time ODM models and WDM models. We will also compare our choice of smooth-k parameters with previous works.

We first note the scaling at small masses of the smooth-k HMF in the case of DM models featuring a cutoff in the power spectrum (approximating equation.\ref{eq:PS_General_Expression}): 
\begin{equation}
\frac{dn}{d\ln{M}}\sim M^{(\beta-3)/3} 
\end{equation}
Hence, $\beta$ controls at the same time the degree of smoothing in ODM models, and the slope of the mass function at small scales in models featuring a suppression of power. 

We show in Fig. \ref{fig:Comparison_Smooth_K_param_Final} the mass functions of halos in WDM (with $m_{\rm th} = 0.25$ keV) at $z=0$ and ODM3 (with $f_{\rm IDM}=1$) at $z=2$. Note that we did not remove artifacts here in WDM. The data points come from two simulations with $1024^{3}$ particles and with box sizes specified on the legends. We also show the sharp-k and top-hat mass functions, along with three smooth-k mass functions with different parameters; the ones from \citealt{Leo_smooth_k}, the ones from \citealt{OtherfitsmoothkODM} and ours. Note that, the smooth-k filter has been used in several other studies to fit the mass functions of DM models featuring oscillations in the power spectrum (see e.g. \citealt{ETHOS_67, 2101.08790}). 

\citet{Leo_smooth_k} fitted WDM models with $(\beta,c)=(4.8,3.3)$. This is indeed the best option to fit our WDM HMF with a smooth-k filter (see red solid line in Fig. \ref{fig:Comparison_Smooth_K_param_Final}). \citet{OtherfitsmoothkODM} matched ODM simulations with $(\beta=3.7, c=3.5)$. We found that a slightly smaller value of $\beta = 3$ (blue solid line) was necessary to fit the smoothed oscillation (see bottom right panel of Fig. \ref{fig:Comparison_Smooth_K_param_Final}). 

Fig. \ref{fig:Comparison_Smooth_K_param_Final} shows that there is not a single value of $\beta$ suited for WDM and ODM simultaneously. Increasing $\beta$ increases the small scale slope in WDM, such that the smooth-k prediction approaches the sharp-k mass function, but reduces the amount of smoothing in ODM (red line right panel of Fig. \ref{fig:Comparison_Smooth_K_param_Final}). It is hence impossible to find a universal set of smooth-k parameters suited simultaneously for every dark matter models.


\bsp	
\label{lastpage}
\end{document}